# HISTCOMP : BIBLIOGRAPHIC ANALYSIS AND VISUALIZATION OF THE "BIOLOGICAL BULLETIN"

**Enrique Wulff**
Marine Sciences Institute from Andalusia (CSIC)
Cádiz, Spain

**Abstract**
A collection of citation data, the HistComp, is available from the Internet as a database of examples of real life citation networks. The purposes of this approach is the analysis of these citation networks on learned literature by presenting its typical steps and results. We have selected the bibliographic insights into the "The Biological Bulletin", the journal published since 1897 by the Woods Hole Marine Biological Laboratory. Since the bibliographic networks tend to be very scattered, their visualization requires of criteria of convergence. To simplify, the main features in such a structure should include the survey for authoritative sources in the hyperlinked environment and the identification of thematic areas. By avoiding excessive loose connections and too dense clustered layouts to be useful, a smooth presentation is obtained by graphically depicting the citation patterns. HistComp computes 8884 articles published by 'The Biological Bulletin' between 1945-2003. A two-dimensional positioning of these papers that represent the extent of their bibliographic coupling and co-citation is offered as a histograph. The criteria to construct it is the adequateness of the visualization relative to the 8884 data set. The spatial representation obtained optimizes the identification of the clusters or topic areas. The thematic importance of marine science involves its participation in 7 of the 7 presenting clusters. The mainstream subjects were crustaceans and echinoderms, with some 60% of the material presented in the graph. But sea anemone, with about 16% of the total, remains as the best visualized topical area. A perspective of the highly relevant papers is readily confirmed by the visual inspection of width of the glyphs used for nodes representation. For user interaction, HistComp employs mouse-over labels.

**Introduction.-**
The basis in this study will be a total of 8884 'The Biological Bulletin' papers from 1945 to 2003, including full reference lists and citation counts to each paper by August 27, 2003.

The *Bulletin* was established in 1897, it is now in its 208th volume, and it is among the oldest peer-reviewed biological publications in the United States. The *Bulletin* is also

59

among the most precious assets of its publisher, the Marine Biological Laboratory (MBL) in Woods Hole, Massachusetts. The MBL is America's oldest private marine laboratory. (Greenberg MJ, 1999)

The initial software for the automatic generation of historical or genealogical maps of papers or topics by processing the export files of source records extracted from SCI was called "histcomp". And an index of HistCite analysis is available under a directory with this initial name "histcomp" (http://garfield.library.upenn.edu/histcomp/) and this bibliomining technique has its own bibliography of papers (http://garfield.library.upenn.edu/algorithmichistoriographyhistcite.html). The software "histcomp" was referred at the Lazerow Lecture to honor the memory of Professor Casimir Borkowski in September 19, 2001, at the University of Pittsburgh.

Nowadays histcomp has evolved into the HistCite[TM] software. This is a system for the historiographic analysis that organizes the bibliographic collections generated by searching in the Science Citation Index of the Web of Science (WOS) or in the SCI-CD-Rom. It permits to follow the evolution of articles, authors, and journals and the graphical representation of the more influential articles on a subject chronology. On November 18, 2002, this program presented at the 65[th] annual conference of the American Society for Information Science & Technology (ASIST) along the bibliometrics session. The authors were E. Garfield (emeritus president of Thomson ISI), A.I. Pudovkin (biologist at the Institute of Marine Biology, Vladivostok) and V.S. Istomin (formerly at Washington State University, now in Vladivostok). The present contribution is based in its 2005 version.

The need of reference librarians and users to improve the results of their searches in databases like SCI, Medline or Chemical Abstracts are well-satisfied by using the "histcomp". The resulting visualization provide a fairly comprehensive snapshot of the "Biological Bulletin". (Boyack KW, 2004)

**Methods.-**
All the references for the 'Biological Bulletin' have been downloaded from the ISI Web of Knowledge, between 1945 and 2003, by using the expression 'SO = (Biological Bulletin)'. With the software HistCite[TM] (2005 version) this set of papers has been graphically represented in a citation network.

The histcomp for the "Biological Bulletin" has been accessed (http://garfield.library.upenn.edu/histcomp/bio-bulletin_all-src/). The main presentation provides sort results by node, author and citation counts. The frequency analysis of author (see Table 1) and journal (not reproduced here because in this case it is limited to the only *Bulletin*, available from: http://garfield.library.upenn.edu/histcomp/bio-bulletin_all-src/hist-jns.html) provided by histcomp is exposed.

Ranked All-Author list.
Total: 8949
Sorted by **pubs**

| # | Name | TGCS | TLCS | Pubs |
|---|------|------|------|------|
| 1 | Atema J | 612 | 122 | 68 |
| 2 | Inoue S | 221 | 17 | 63 |
| 3 | BROWN FA | 782 | 135 | 56 |
| 4 | Valiela I | 127 | 16 | 53 |
| 5 | Zigman S | 70 | 13 | 53 |
| 6 | Barlow RB | 228 | 41 | 51 |
| 7 | STUNKARD HW | 471 | 82 | 51 |
| 8 | KOIDE SS | 108 | 11 | 47 |
| 9 | METZ CB | 310 | 39 | 45 |
| 10 | Armstrong PB | 75 | 15 | 43 |

**Table 1**.- Sequence of the 10 first authors in bibliography on 'The Biological Bulletin', sorted after the number of articles by the authors in this journal. Clicking on the hot linked number under Pubs shows a list of the articles by author.
http://garfield.library.upenn.edu/histcomp/bio-bulletin_all-src/hist-aus-pubs.html

This is supplemented with some data as supplied by HistCite™ (2005 version) concerning the country, document type, institutions, publication year, subject category, and word frequency.

The identity of the core literature is examined by considering the selection threshold used to produce the graph 'LCS > 12', and by implementing the outer references frequency ranked tables and the missing link tables. A combination of both tables will serve to improve the retrieved original information collection.

The citation matrix that histcomp manages permits the elucidation of the line forces guiding the elaboration of the flow chart. It can be used to visualize the co-citations.

**Characterization of the 'Biological Bulletin' between 1945 and 2003. Basic analysis.**
The 'Biological Bulletin' authorship geographic distribution spreads over 62 countries. This journal publishes articles (54%), meeting abstracts (44%), and other kind of editorial material (1%), notes, reviews and letters. After the available data (57.3% of the records do not contain data in this field) some 1034 different institutions are involved with getting published by the journal, and the three first European universities are those from Palermo (Italy), Basel (Switzerland) and Barcelona (Spain). The maximum number of annually published distinct record material was 242 in 1960. This journal is purposely committed 100% with marine & freshwater biology. Some 9042 different authors see their material introduced to the topical structure of the discipline through the pages of the "Biological Bulletin". These are data provided by WOK after its option 'Analyze results' that view rankings and histograms of the authors, journals, etc for each of the retrieved set of records; the data have been checked after the same rankings and histograms as supplied by HistCite[TM].

**The 'Biological Bulletin' after the histograph.-**
Histcomp works on the basis of the model of circles (see the graph at http://garfield.library.upenn.edu/histcomp/bio-bulletin_all-src/graph/1.html, and a fragment at Table 2). The area of each circle is proportional to the number of articles that cites to the one pointed out with a number inside the circle. With the help of a list of authors, "Ranked all-author list" (see, Table 1) and ordered by TLCS (total local citation score), we have a criterion of local citation to the collection. This is the first perspective of the highly relevant papers in the "*Bulletin*". A visual inspection of the width of the glyphs used for nodes representation readily confirms it. For user interaction, histcomp employs mouse-over labels.

The LCS is the local citation frequency inside the collection. It is particularly suited in this case because the graph proposed considers a selection threshold 'LCS>12'. It means that all the articles that display 12 or more citing articles inside the "*Bulletin*" are shown in the graph. Histcomp also works with another type of frequency, the global citation frequency (GCS), GCS, is the global citation score based on the ISI Web of Science (WOK) database record.

The problem of discernment of the citation cycle that exist inside the bibliography displays a first insight depicting the activities of Dr CM Williams from 1946 to 1968 and, Dr DM Skinner from 1962 to 1972. A solid line links node #4208 to node #78. Although virtually absent nowadays, insects are the scope of the published material by Williams and crustaceans were the marine invertebrates reported by Skinner. The common focus of both scientists was the endocrine system.

The second citation cycle (see Table 2) provides a graphical patron for the identification of critical works regarding the topic 'sea anemone'. To aid in the recognition of appropriate evidence, the graphical interface permits the visualization of the particular records corresponding to each node. As extracted after the graph, it is clearly visible from the records that two dissertations (ROBERTS BJ, 1941, THESIS STANFORD U) and (BUCHSBAUM V, 1968, THESIS STANFORD U) are the basis for the work of the papers authored by Dr L Francis, which is the original contributor on the subject. So the first level document tipology is composed by each of the papers considered in the graph. The second level provides the references that constitute the bibliographies of each of this papers. By using the more recent HistCite[TM] the reference librarian can obtain a graphical view limited to the topic 'sea anemone', understanding the history of the research question.

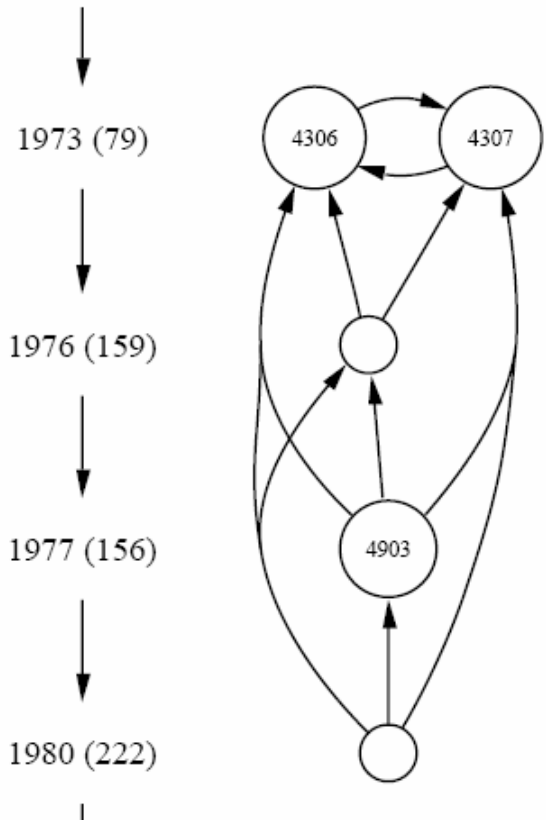

1973 (79)

1976 (159)

1977 (156)

1980 (222)

**Table 2**.- 'The Biological Bulletin' second citation cycle, provided by its graph when the number of citing nodes inside the journal is over 12 (LCS > 12). It describes the topic 'sea anemone'. The circle represent papers, and the number inside the circle is the node number. http://garfield.library.upenn.edu/histcomp/bio-bulletin_all-src/graph/1.html

The remaining five clusters are indicative of simple hierarchical choices limited to the visual exploration of a couple of cited/citing papers. So graphically the depicted relation is simple. Although when the second level document topology is used it becomes possible to present the extensive ramification of the continuous flow of contributions inside '*The Biological Bulletin*' that communicate with the original paper on the topic and through bibliographical coupling. In these five cases a topic structure has been derived by matching of the provided 'histcomp' with 'HistCiteTM' graph. The animals studied were clams, asteria, oyster (on eggs fertilization and larvae breeding), crustaceans, intertidal molluscs, lobsters, gastropoda and echinoiderms. Self citations dominate the citation statistics, percentile rankings going from 100% to 33%.

**The Citation Matrix.-**
A stage in the visualization of "The Biological Bulletin" corpus is considering the citation matrix that 'HistComp' offers (see Table 3).

Articles from *Biological Bulletin*, 1945-2003 (Wed Aug 27 09:51:48 2003)
Nodes: 8884
Sorted by **year, journal, volume, page**.
Page **1**:  1  2  3  4  5  6  7  8  9  10  11  12  13  14  15  16  17  18

| cited nodes | Cited nodes | Nodes | GCS | LCS | citing nodes |
|---|---|---|---|---|---|
| | 0 | 1 1945 VONBONDE C | 3 | 1 | 17 |
| 1397 | 1 | 4301 1972 ZEUTHEN E | 5 | 0 | |
| 3309 | 1 | 4302 1973 ATWOOD DG | 20 | 6 | 4547 4845 5007 5810 7143 7534 |
| 1870 | 1 | 4303 1973 BRITZ SJ | 7 | 0 | |
| 3429 | 1 | 4304 1973 BUCK J | 21 | 3 | 4581 4842 5169 |
| 3452 3483 3874 | 3 | 4305 1973 ELDER HY | 33 | 1 | 4418 |
| 4307 | 1 | 4306 1973 FRANCIS L | 111 | 22 | 4307 4538 4717 4840 4903 5002 5214 5377 5380 5610 5746 5782 6196 6208 6213 6764 6766 6782 6956 7292 7412 8731 |
| 4306 | 1 | 4307 1973 FRANCIS L | 140 | 31 | 4306 4717 4840 4842 4903 5002 5214 5377 5380 5610 5746 5782 5948 6003 6142 6184 6196 6213 6405 6764 6766 6782 6941 6956 6987 7065 7188 7217 7292 8069 8731 |
| | 0 | 4308 1973 FRANZ DR | 14 | 3 | 6532 7608 8444 |
| | 0 | 4309 1973 FRIESEN LJ | 19 | 0 | |

**Table 3**.- Citation matrix for the journal 'The Biological Bulletin' (partial view).
http://garfield.library.upenn.edu/histcomp/bio-bulletin_all-src/index-cm.html

64

Like a very torn and deformable fishing net (Price DJD, 1986. p. 268) is the structure of alternating cited and citing nodes. The pattern of linkage represents an item, its number of cited nodes, and the numerical codes identifying them, its global citation score (GCS) and its local citation frequency (LCS). The last data is set of citing nodes. Co citations and bibliographic couplings are at reach from the citation matrix. Any couple of cited nodes are a co citation, and all the pairs of citing nodes are bibliographically coupled. The matrix permits clustering using citing nodes (bibliographic coupling groups documents) and clustering using cited nodes (co citation links documents) (Morris SA et al., 2003).

The matrix leads to the identification of co citations easily viewed with the graph. For example, nodes 4306 and 4307 are co cited by nodes 4717, 4903 and 5380. Their interrelation is special for the 'sea anemone' cluster.

This relational structure can be called a "subject space" (Price DJD, 1986). It is said that this structure provides a natural and automatic "indexing". So built into the network linkage of the entire collection of "The Biological Bulletin" is a structural scheme. And we can traverse this map by using the citation matrix.

The idea behind understanding the informative effects of this citation matrix is the concept of additivity of the levels of reference. It means that inside the journal "The Biological Bulletin" an author that must only refer in his publications to his reading domain (the universe of articles that an author has read to write a text) is said to be in the zero reference level. He will be part of his one reference level if he is able to include in his references' lists the publications that were grouped in the zero level, plus the material that was referred inside the bibliographies of the zero level, and so on. We use the concept of additivity by assimilating it to linearity (a magnitude depends on others which are the result of a sum (Shapley Ll S, 1953). Therefore, the activity in local networks (the whole source papers published by 'The Biological Bulletin') of the authors finds a model that follows the pattern of the analysis of references (Kessler MM, 1966).

The article at node Nº 4555 (authored by Webster SK in 1975, see "Nodes" in http://garfield.library.upenn.edu/histcomp/bio-bulletin_all-src/index-10.html) includes in its list of references 4 articles (Nº 1246, 3281, 3342, 4167) from the local collection (as identified by the column "Cited nodes", that expresses the local cited references or papers published by the 'Biological Bulletin' and employed as reference by Webster). SK Webster has read these articles; they are part of his reading domain, of his zero reference level. But, particularly, the article Nº 4167 is also part of his one reference level, because it can be admitted that Dr Webster has read the work of Dr K Johansen (art. Nº 3342) after finding it between those selected by Dr RJ Ulbricht (art. Nº 4167). And again the article Nº 3342 provides the two reference level as Dr Webster could have read the paper of Dr AC Giese (art. Nº 3281) once having located it in the list of reference of Dr K Johansen (art. Nº 3342).

**Procedures for improvement and correction of the output of literature searches. The 'Outer references' and the 'Missing links'.**

If the percentual selection threshold has been well chosen (here LCS > 12) then the user will obtain core papers of prime interest. But 'histcomp' produces a list of highly cited works outside this initial bibliography, the outer references. (see Table 4)

ISI Web of Science location: [                    ]

Cited references outside of this network.
Total: 80764 (top 300 shown).
Sorted by **LCS**.

| # | **LCS** | **Reference** |
|---|---|---|
| 1 | 103 | LOWRY OH, 1951, J BIOL CHEM, V193, P265  WoS |
| 2 | 64 | SOKAL RR, 1981, BIOMETRY,  WoS |
| 3 | 53 | LAEMMLI UK, 1970, NATURE, V227, P680  WoS |
| 4 | 51 | BRADFORD MM, 1976, ANAL BIOCHEM, V72, P248  WoS |
| 5 | 51 | THORSON G, 1946, MEDD KOMM DAN FISK P, V4, P1  WoS |
| 136 | 11 | KLEINHOLZ LH, 1936, BIOL BULL, V70, P159 WoS |
| 144 | 11 | LILLIE FR, 1915, BIOL BULL, V28, P22 WoS |
| 216 | 11 | SCHARRER B, 1944, BIOL BULL, V87, P242  WoS |
| 267 | 12 | TYLER A, 1941, BIOL BULL, V81, P190  WoS |
| 291 | 11 | WILSON EB, 1903, BIOL BULL, V4, P197  WoS |

**Table 4**.- 'Outer References' – Top five nodes exterior to the original bibliography and five 'The Biological Bulletin' reference outside the 'histcomp'.
http://garfield.library.upenn.edu/histcomp/bio-bulletin_all-src/out-refs.html

This list is identified by the software 'histcomp' sorting the references used by the original core collection by citation score. These could be not only articles, but books and patents. For those included in WOS, a hotlink is provided which leads to the WOS search engine. The librarian can decide (Garfield E, Pudovkin AI & Istomin VI, 2003) whether to add these candidate references to the bibliography. For "*Biol. Bull.*" 5 articles do not turn up in the original WOS search (out of the 80764 (top 300 shown) cited references outside of this network listed by 'histcomp' as outer references). They are all previous to 1945, so they were not considered as source items at the moment of retrieval.

The WOK workaday makes conceivable the commission of mistakes in the articles introduction. Errors can be corrected and the routine for correction in 'histcomp' is called 'Missing Links' (see Table 5). It reports on the potential bibliographic description missed data, by checking every doubtful reference against the main file. 116 were potentially missed citations in the 'Biological Bulletin' histcomp. Nodes have citations that may potentially refer to other nodes because of inconsistencies on pagination, inconsistencies of hyphenation, introduction of non-standard expressions in the bibliographic quote, etc.. These were the most frequent problems identified and it illustrates the usability of this device in the case of the 'Biological Bulletin'.

Potentially missed citations...
116 nodes have citations that may potentially refer to other nodes.
1 | 20 1945 BIOLOGICAL BULLETIN 88(3):254-268
SPIEGELMAN S; STEINBACH HB
*SUBSTRATE-ENZYME ORIENTATION DURING EMBRYONIC DEVELOPMENT*

---

`SPIEGELMAN S, 1945, UNPUB BIOL B, V89, may refer to` 28
`SPIEGELMAN-S-1945-V89-P122`
2 | 173 1947 BIOLOGICAL BULLETIN 92(2):115-150
LYNCH WF
*THE BEHAVIOR AND METAMORPHOSIS OF THE LARVA OF BUGULA-NERITINA (LINNAEUS) - EXPERIMENTAL MODIFICATION OF THE LENGTH OF THE FREE-SWIMMING PERIOD AND THE RESPONSES OF THE LARVAE TO LIGHT AND GRAVITY*

---

`MILLER MA, 1946, BIOL B, V90, P121 may refer to` 69 `MILLER-MA-1946-V90-P122`

**Table 5**.- 'Missing Links' : Potentially missing citations, and 'variations'.
http://garfield.library.upenn.edu/histcomp/bio-bulletin_all-src/miss-links.html

**Conclusion**
From the bibliographical instruction (McInnis R, 1982) point of view those unfamiliar with a topic have a first and foremost information space with the histcomp for "The Biological Bulletin", published at Woods Hole. By identifying the main authors and considering the indicative labels provided by the graph.

As a frame of reference 'histcomp' is also useful for retracing the history of any research question, and it permits to the collection development librarian to orient the purchases by considering the main journals by looking at the importance of the articles.